# AI VIRTUE: WHAT IS "GOOD" KNOWLEDGE: IN THE AGE OF ARTIFICIAL INTELLIGENCE?

Alan Liu
University of California, Santa Barbara

**ABSTRACT**
In the age of AI, what will be good knowledge? This article, which is accepted and forthcoming in a special issue of *Modern Fiction Studies* on "Cultural AI" in 2027, applies digital humanities methods to map epistemic virtues (like "true," "accurate," "creative") used in a corpus of 553 journal articles on AI published in 2024. "Creativity" comes in for special attention as an example. Exploring this discourse of value, the article considers how a framework might be developed for evaluating the knowledge-worth of AI—one less locked into values formed around pre-AI "knowledge work" agents or structures, and more open to the future values of "generativity." The essay is supported by an online digital kit for exploring data models of the corpus of articles on AI it studies.

In the age of generative AI, what will be *good* knowledge?

This epistemic question branches into myriad inquiries about specific knowledge values enhanced or impeded by AI—for example, truth, accuracy, or transparency. Related are qualities that overlap with, but are not strictly bound to, knowledge, including ethics and creativity. Creativity will be my case study here. Ultimately, questions about AI and good knowledge extend to social, political, and economic issues bearing on the fact that AI is a late chapter in our contemporary era of "knowledge work" (the subject of my book, *The Laws of Cool: Knowledge Work and the Culture of Information*). Questions that arise in this context include, *who* will be rewarded and held accountable for using AI to work with knowledge? And which organizational, governmental, and other institutions will do the rewarding and accounting?

Time will be needed for answers to these questions to emerge and, of more consequence, to become norms. After all, every new knowledge technology—inseparable in the last instance from media, information, and communication technologies (writing, print, TV, the Internet, and others)—debuted to the same normless mix of hype and skepticism that greeted ChatGPT. We remember that Socrates, the oral philosopher, dismissed writing on epistemic grounds. In the *Phaedrus*, Plato's Socrates tells of the god Theuth who demoed for the Egyptian king his new invention, writing: "This, said Theuth, will make the Egyptians wiser and give them better memories; it is a specific [*pharmakon*] both for the memory and for the wit." The king responded:

> O most ingenious Theuth, the parent or inventor of an art is not always the best judge of the utility or inutility of his own inventions…. [F]or this discovery of yours will create forgetfulness in the learners' souls, because they will not use their memories; they will trust to the external written characters and not remember of themselves. The specific which you have discovered is an aid not to memory, but to reminiscence, and you give your disciples not truth, but only the semblance of truth; they will be hearers of many things and will have learned nothing; they will appear to be omniscient and will generally know nothing; they will be tiresome company, having the show of wisdom without the reality (Plato).

How easy it would be today to echo: *O ChatGPT! You give learners not truth, but only the semblance of truth; they will be hearers of many things and will have learned nothing; they will appear to be omniscient and will generally know nothing; they will have the show of wisdom without the reality.*[1]

[1] Cf., Dan Rockmore's use of the same passage from Plato to discuss large language models.

Centuries had to pass in early-literate societies, as Michael Clanchy shows in *From Memory to Written* Record*: England, 1066-1307*, before practices, institutions, and laws could be laid down to allow people to trust writing. Only after such normalization could writing be judged not just as good or bad memory (the consummate epistemic value in oral cultures where people must, as Socrates put it, "remember of themselves") but on the basis of knowledge values that altered the ideals of memory itself—e.g., in the direction of reproducibility, precision, and, as a post-Romantic add-on, once external media seemed to secure memory, creativity (as in Wordsworth's creative memory in *The Prelude* or Nabokov's in *Speak, Memory*).[2]

Of course, in accelerative late modernity it will not take centuries to normalize AI. Yet as Arvind Narayanan and Sayash Kapoor argue in their important essay, "AI as Normal Technology," the process will not be instantaneous either. Effective operationalization of AI will likely lag its innovation by a decade or more in the manner of past tech rollouts such as those of electricity or computers (Narayanan and Kapoor 3–12). And it is not just time that will tell. Equally crucial will be whether a sufficiently expansive, whole-society framework of evaluation *capable* of forming norms can develop—i.e., a common intellectual, cultural, and civic framework for appraising AI across domains and scales of address. *Norms* in this formulation is related to what Colin Milburn and Rita Raley, in a recent critique, call "appropriate" (or, objecting to appropriative large language models, ideally "inappropriate") AI-use policies now being improvised by, among other institutions, universities.[3] *Domains* means differentiated social sectors, institutions, and fields. And *scales* refers to stepped levels of entities, ranging from individual users and technologies (as well as agents, processes, and resources in tech itself) to larger production and social units such as work teams, departments, organizations, industries, social groups, and nations.

At present, we tend to evaluate AI just by jumping between preset domains and scales—for example between discussing how good or bad AI is for individual users organizations, industries, groups (e.g., students or young people), and the general public.[4] Such domains and scales are preset in the sense that they align with *pre*-AI sociotechnical production and market entities—in knowledge work, for example, employees using databases and networks—and are thus default frames of evaluation. In discussing AI, therefore, we traverse domains and scales in abrupt jumps along presettings—like shuffling a playlist. We assume that the workflows connecting underlying production, consumption, and other structures conform to the *status quo ante*, and do not rethink them in a manner open to new combinatorics of people and technologies changing behaviors in ensemble in a total knowledge space now bending, as it were, around the black hole of AI. As AI warps data, information, and knowledge space-time, however, sociotechnical agents and processes will inevitably migrate, hybridize, and fuse to make, for instance, the now seemingly intuitive assemblage *user + AI + training dataset* as obsolete as *driver + horse + wagon* from an earlier industry.

The *teams + AI* space is at present especially lively with new combinatorics of human-machine agents. How will teams mutate when AI becomes one of the team? It is not just that people will work with AIs in their teams, but that they will also start teaming AIs. Many already use one LLM to improve prompts for, or review results from, another. But as introduced in Anthropic's Claude Opus 4.6 in early 2026, the state of the art evolved to the *designed* teaming of multiple instances of an LLM:

[2] On writing's ability to take on new missions of originality and creativity, see Ong, "Romantic Difference and the Poetics of Technology."

[3] For example, my own University of California system's 2026 *Report of the Academic Senate Workgroup on Artificial Intelligence* advises a "center-periphery" policy balancing a "central framework" against "disciplinary norms."

[4] See the corpus study by Sieber et al. on how the AI research literature speaks of different "publics."

> Agent teams let you coordinate multiple Claude Code instances working together. One session acts as the team lead, coordinating work, assigning tasks, and synthesizing results. Teammates [separate LLM sessions] work independently, each in its own context window, and communicate directly with each other (Anthropic).

LLMs like Claude can also run *internally* as what are essentially teams "swarming" with parallel and sequenced subagents.[5] Then, on top of all that, open or third-party frameworks like Paperclip orchestrate *different* LLMs as agent teams, allowing for the start-up of whole "tiny team" companies staffed by one or a few humans and multiple AIs simulating an entire team organization.[6]

All this raises the prospect of a McLuhan "medium is the message" effect. As multi-agent AIs propagate, their internal technically-optimized team protocols will increasingly reshape the behavior and psyche of the human teams using them, altering how humans act as social beings in organizations. Such may lead to what Vladimir Menshikov et al., applying Niklas Luhmann's sociology to AI, call an "artificial sociality" of "communication between [AI] agents operating autonomously in a self-organising network that is autopoietic"(217). The AI research community, it may be noted, has already begun exploring such artificial sociality as part of the *teams + AI* problem space. The Hybrid Human Artificial Intelligence international conference in 2025, for example, included sessions such as "Advancing Human-Machine Teaming" (Verhagen et al.).

In the 1990s, as I note in my *The Laws of Cool*, the buzz terms for postindustrial neo- and flex-forms organizations were *flat*, *lean*, *disintermediated*, *virtual*, and *networked*. Today the buzz is all about how AI will remake companies into *agentic organizations* with *matrix*, *lattice*, *fractal*, and *holarchical* (holistic rather than hierarchical) structures powered by *agent-led teams*, *agent factories*, and *AI-first cross-function pods*—all braided together with LLMs.[7] It is almost as if the embedding spaces, neural networks, and transformer processes of large language models will now simply *be* the new theory of organizations and, at the level of individuals, of humans working in such organizations. Consequently, the hype is now also about how AI will mutate the fundamental authority structures of organizations (e.g., as plotted on the Hackman authority matrix for teams[8]) while also, as the McKinsey consulting firm speculates, mutating the core business, operating, governance, workforce culture, and technology and data models of tomorrow's organizations (Sukharevsky et al.).

Mutant buzz, biz, and hype-speak like this, however, is not yet the norm. Instead, to return to my main argument, discussions of AI at present normally adhere to preset domains and scales of address. Thus, consider how some of the best general-audience writings discuss AI in a public-intellectual, memoirist, or similar voice. The scale problem here is especially noticeable. Often the fulcrum in newspaper or magazine *pensées* about AI is the individual user's experience of specific tools like chatbots, meaning that generalizing about larger social bodies and technological methods is a stretch. A rich, thoughtful example is Meghan O'Rourke's 2025 opinion piece in *The New York Times* titled "I Teach Creative Writing. This is What A.I. Is Doing to Students." Toward her conclusion, O'Rourke reflects,

> As a poet, I have shaped my life around the belief that language is our most human inheritance: the space of richly articulated perception, where thought and emotion meet.

[5] On swarm parallelism and subagents in Claude Code, see Franzen.

[6] On "tiny teams," see Griffith; Bentes; and Kelley.

[7] See e.g., Minnaar; Janse; D. O'Reilly; and Henkel et al. on some of these new organizational structures. For forecasts of how AI will drive such structures, see e.g. Sukharevsky et al.; Voccola; and Williamson and Lanz.

[8] For the Hackman Authority Matrix, see Hackman 52. For an example of its use in forecasting how AI will reshape organizations, see Williamson and Lanz.

> Writing for me has always been both expressive and formative—and in a strange way, pleasurable.
>
> I've spent decades writing and editing; I know the feeling—of reward and hard-won clarity—that writing produces for me. But if you never build those muscles, will you grasp what's missing when an L.L.M. delivers a chirpy but shallow reply? What happens to students who've never experienced the reward of pressing toward an elusive thought that yields itself in clear syntax?
>
> This, I think, is the urgent question…. For [today's generation growing up with A.I.], the chatbot won't be a tool to discover … but part of the operating system itself. And that shift, from novelty to norm, is the profound transformation we're only beginning to grapple with. (O'Rourke)

Here the proving ground for experiencing and evaluating AI is a philosophically wise, meditative, and eloquent *I* and *me* that projects its subjectivity through the delivery mechanism of the pronoun *you* ("if you never build those muscles, will you grasp what's missing…?") onto students constituted as subjectivities-in-training. The aura of this personal voice is so strong that even when O'Rourke's discourse unfolds outward to embrace collective personae ("our most human inheritance," "we're only beginning") it is difficult to overcome the impression that an entire student generation, their educational institution, and their society are anything other than aggregates of personal sensibility. O'Rourke values the individual's experience of struggling "toward an elusive thought," and so devalues the "norm" (operationalized as standardization) that will come when AI becomes everyone's "operating system" for it all. But it may be that such an evaluation cannot persuasively be made without integrating perspectives across scales, which would change the problem of individuation versus standardization from an either/or crux into the *balancing* of "me" against shared norms constitutive of any society, and of the mediated and regulated processes for such balancing constitutive of modern societies.

The scholarship on AI, of course, is typically less personal in voice than public-intellectual writings like O'Rourke's. But scholarship too is flummoxed by impasses between domains and scales. My sample consists of 553 journal articles published in English in 2024 that mention "artificial intelligence" or "AI" in their titles. The sample includes the 227 "highly cited" such articles of that year in the Web of Science Core Collection index; and the 326 articles mentioning AI in their titles in the Web of Science Arts & Humanities Citation Index (AHCI). (Hereafter, I refer to these article sets as my WoS Core and WoS AHCI corpora.)[9]

---

[9] I conducted my search of the Web of Science for journal articles on Feb. 15, 2025. See Folder 1's "readme_folder_1.docx" in my online "Kit for Exploring Articles on AI Published in 2024" for details on my search terms and filters, and for details about how I preprocessed article texts. The following goals and constraints influenced my corpus design. Because my main concern is what I term "norms" in understanding and using AI, I searched widely in Web of Science for both highly-cited articles in its Core Collection citation index as the basis of one corpus and, separately, for relevant articles in its AHCI (arts and humanities) citation index as the basis for another corpus. I searched AHCI separately, though it is a subset of the Core Collection index, because otherwise there would have been no representation of articles from arts and humanities journals at all, which I needed because my case study concerns creativity. (At the time I searched, "highly cited" articles for 2024 in WoS Core included not a single AHCI article, though ten appeared later.) Among constraints on corpus design, some of the most important are the following. I relied on citation indexes from Web of Science (previously called Web of Knowledge) because its main competitor, Scopus, is an Elsevier product not available through my institution. I limited myself to articles published in 2024, and to those with "artificial intelligence" or "AI" in their titles, because I had to manually download and preprocess articles to create my corpora. 553 articles were at a scale I could accomplish. Also, I limited myself to searching only for articles in Web of Science's indexes of *published* articles, and not also, or instead, for preprints in Web of Science's Preprint Citation Index. This is also due to the need for disciplinary breadth in studying norms. Leading-edge research on AI in technical fields, or by technically-oriented

The impasse I indicate can be glimpsed by looking at representative articles on major topics identified through a topic model I made of my WoS Core corpus (optimized at 22 topics). The model is dominated by topics about healthcare, business, education, and regulatory or public policy. For example, topics include (in labels I assigned with the help of AI itself), "AI in Healthcare and Clinical Practice," "Medical Imaging and Diagnostics," "AI in Workplace and HR Management," "AI in Education and Academic Writing," and "AI and Policy."[10] Using the model as guide, I read a sampling of articles strongly associated statistically with each topic to observe what their authors focus on. As might be expected, researchers think about individual AI-users in their organizational roles as employees, consumers, doctors, patients, students, and so on. But they also put the spotlight on AI tools. And, of course, the future of their organizational, institutional, or disciplinary sector is a framing concern. The result is discourse that jumps among preset domains and scales with sparse conceptual or rhetorical means for coordinating them—or more radically, for questioning if the very nature of such domains and scales might not be provoked by AI into new configurations.

Consider, for instance, an article by Wenxiu Li et al. associated with the "Medical Imaging and Diagnostics" topic in my WoS Core corpus. The authors propose an AI-assisted breast cancer diagnosis system for developing nations. At times, they foreground the perspective of individual doctors and patients—e.g., "doctors need to see a large number of patients with similar symptoms every day…. Furthermore, if patients do not receive efficient health care [it] may aggravate, damage their health, and increase their financial burden" (Li et al. 396). But they often override that perspective with an impersonal *we* and *it* voice speaking for the technology of the proposed AI system, or for the medical institutions in which it is to be embedded—e.g., "Based on hospital electronic health records, we propose a breast cancer staging-assisted model. The BC-INIT-CNN model used a novel condition knowledge representation method to integrate the condition knowledge into the neural network model" (411). Like so many others in my corpus, the article jumps discontinuously in montage cuts between disparate points of view and subjects because it silently assumes that a stable epistemic and sociotechnical framework holds everything together.

Similarly, a topic model I created of my WoS AHCI corpus (coincidentally also optimal at 22 topics) identifies topics in the humanities and arts related to writing education, literary studies, philosophy, music, religious studies, and others.[11] Reading articles associated with these topics shows that the humanities and arts focus on individual authors, teachers, students, or (in psychological or phenomenological approaches) consciousnesses. But they also focus on the impact of AI on their disciplines, and on the wider sociocultural scene. Again, the result is a discourse poorly equipped to

---

scholars in other fields, often first circulates in preprints in arXiv.org and other preprint repositories. But many other fields do not publish preprints. It would be a separate project to sample the preprint literature, or to create a corpus combining published articles and preprints. (The latter is difficult due to the lack of a "highly cited" ranking common to Web of Science's article and preprint indexes. One would need to invent a rationale for sampling both kinds of documents based on a standard criterion or, failing that, some standard of representativeness or statistical normalization.) The different goal of such a separate study, for example, might be to study the divergence or lag between general scholarly discourse and advanced or pioneering discourse on AI. (Thanks to Richard Jean So for pointing out that preprints would be a valuable alternative corpus.)

[10] I created my topic model using the Mallet tool (McCallum). For details on my modeling, optimization, and topic labeling process, see "readme_folder2.docx" in the digital kit accompanying this essay (Liu, "Kit for Exploring Articles on AI Published in 2024", folder 2). The full set of topics in my 22-topic model of WoS Core is summarized with labels in a spreadsheet summary in the kit's folder 2.

[11] The topics in my 22-topic model of WoS AHCI are labeled and summarized in a spreadsheet in Liu, "Kit for Exploring Articles on AI Published in 2024", folder 2. The kit also includes a 40-topic topic model of WoS AHCI I used. The latter is not as optimal for coherence, document entropy, and other factors. But in my judgement its finer-grained topics are more interpretable.

evaluate AI across domains and scales. An example is an innovative article by Amy Stornaiuolo et al. on "Digital Writing with AI Platforms: The Role of Fun With/In Generative AI." The article studies an aspect of AI in education rarely discussed amid all the doom and gloom about cheating, bias, and other pathologies: *fun*. An incentive for students to use AI, the authors find (through surveys and focus groups) is doing it "just for fun," "messing around," and being "playful and creative" (Stornaiuolo et al. 84). However, the perspective also soars to that of critiquing the largest, least fun aspects of AI:

> Such an understanding of how humans and tools are coconstitutive along social, political, economic and environmental dimensions is central to what some are calling the "postdigital" turn in education…. [P]ostdigital approaches highlight issues of social power that are embedded in platform architecture … [including, citing another researcher] social arrangements of racial technocapitalism (85).

While the article is persuasive that both the student's and larger social viewpoint are needed, it is difficult to make the stretch from student fun to the critique of "social, political, economic and environmental dimensions" and "racial technocapitalism."

My opening question "what will be *good* knowledge?" thus ultimately discloses my more general question: what can be an adequately expansive, integrated framework for judging the value of AI? I do not pretend to have the whole-society answer. But I have suggestions for the intellectual part of the framework, acknowledging that many other perspectives—e.g., technical benchmarking, stock-market performance, and media vibe—will weigh in too.

One possible intellectual framework is ontological, *à la* approaches that decenter the human in philosophy, science technology studies, large technical systems theory, media archaeology, and other areas of the contemporary humanities and social sciences. Often poststructuralist in flavor, such approaches erode, even as they re-create, the values once identified as human. Well-known thinkers in this mode include Michel Foucault on "What Is an Author?", Gilles Deleuze and Félix Guattari on "assemblages" and "bodies without organs," Bruno Latour on actor network theory, Thomas P. Hughes on large technical systems, Donna Haraway on cyborgs, Bernard Stiegler on "technics," N. Katherine Hayles on "cognitive assemblages" of "humans and technical systems," and Benjamin H. Bratton on the "Stack."[12] The main idea is that agentic assemblages, from the earliest human with a tool to the latest with AI, are not fixed *a priori* but instead emerge, merge, and crisscross in domain and scale through what Deleuze and Guattari call "lines of flight" and actor-network theory terms "detours," "translations," "delegations," or "compositions."[13]

An article from my WoS Core corpus applying this approach is "Academic Communication with AI-Powered Language Tools in Higher Education: From a Post-Humanist Perspective" (Ou et al.). The authors "demythologise AI" by describing it as "distributed cognition, spatial repertoire, and assemblage" and "an expansive and polycentric framework of communicative repertoire." Their goal is to grasp AI as a "co-agency of human and non-humans … in shaping language practices and literacy" (Ou et al. 3).

---

[12] Foucault; Deleuze and Guattari; Latour, *Reassembling the Social*; Hughes; Haraway; Stiegler; Hayles, esp. 115-141; Bratton. I include Hughes (and large technical systems theory) in this grouping even though Hughes is not poststructuralist in style because his theory fundamentally aligns with poststructuralist arguments, as in the case of his integration of "reverse salients" in the development of large technical systems (79-105).

[13] See, e.g., Deleuze and Guattari 3; and Latour, "On Technical Mediation: Philosophy, Sociology, Genealogy" and "An Attempt at a 'Compositionist Manifesto'."

Setting aside the ontological framework, however, I propose a less familiar epistemological framework more specifically suited to evaluating AI . I suggest using what philosophers call "virtue epistemology" to explore "epistemic virtues" and "epistemic values" circulating in the discourse about AI, and around, AI.[14] *Epistemic virtues* refers to values we endorse whenever we characterize people as good knowers. They are open-minded, impartial, brave, and so on. By contrast, *epistemic values* refers to values such as truth, accuracy, or transparency that primarily characterize the ideas, theories, forms, and practices of knowledge itself. In reality, though, it is hard to maintain the divide between such virtues and values. Terms such as *coherent*, *rigorous*, or *creative*, for instance, describe both the knower and the known. I will here thus simplify by using *epistemic values* to mean both virtues and values, separating terms only as needed.

A few preparatory observations about epistemic values will be helpful. One is that to my knowledge there is no complete, or even very large, declared set of such values.[15] Instead, the philosophical literature relies on short, partial lists that in practice are *a posteriori*, whatever their relation to the philosophically *a priori*. Epistemic values manifest when researchers speak them, and they assume the formal status of values when philosophers notice them.[16] Here for instance is a representative list in Rik Peels's "Epistemic Values in the Humanities and in the Sciences":

- academic integrity;
- coherence;
- empirical adequacy;
- explanatory truth;
- insight;
- intellectual thoroughness;
- justified belief;
- knowledge;
- making sense of things;
- open-mindedness;
- rational belief;
- responsible belief formation;
- tradition and tradition transmission;
- true belief;
- trust among peers;
- understanding;
- wisdom. (Peels 98)

Secondly, epistemic values are ill-sorted in their word forms and inflections. It is unpredictable which member of what applied linguists call a "word family" does, or does not, connote epistemic value.[17] For example, some words are recognized as epistemic values in both noun and modifier forms (e.g., *truth* and *true*), while others are recognized as values primarily in one form or the other (e.g., *systematic* as opposed to *system*). Yet this very linguistic heterogeneity is of interest. Epistemic values are linguistic cosplayers dressing up in different idiomatic forms to suit themselves to motley subjects, objects, and actions, and in some cases to transfer functions and attributes between those in a masquerade dance of truth. Values originally suiting one kind of being (human) are thus now busily readdressing AI, allowing us to say, for example, that ChatGPT is, or is not, *true*, *coherent*, or *creative*.

[14] The following is a good initial reading list on epistemic virtues and values in "virtue epistemology" philosophy: Peels; Dongen and Paul; Turri et al.; Schindler; and Fricker.

[15] Whether there is a *canonical* list is a separate question that to my knowledge has not been addressed.

[16] See Schindler for a quantitative study (based on a questionnaire) comparing the epistemic values recognized by scientists and philosophers.

[17] See Bauer and Nation on "word families." The word family in the BNC/COCA basewords model, ver. 1.3.0, for the headword *system*, for instance, is *system*, *systems*, *systematic*, *systematically*, *subsystem*, *subsystems*, *unsystematic*, *unsystematically* (Anthony, *WordFamilyFinder, Ver. 1.0.1*).

Thirdly, while epistemic values are in principle distinct from other values (Peels: "epistemic … values ought to be distinguished from other kinds of values, such as aesthetic, moral, economic, prudential, social, and political values," 97), they are in fact rarely thus immaculate. After all, that is why we set store by them *as* values in the first place. The worth of knowledge is negotiated in a trader's lingua franca equating its value with other social, economic, political, cultural, and other goods.

Finally, while there is no complete list of epistemic values, even a partial list quickly suggests *structures* of value. For example, in figure 1 I hypothesize a plausible structure of just a few epistemic values that oppose rationalist and non-rationalist knowledge.

| **Rationalist Epistemic Values** | | |
|---|---|---|
| Accurate | Objective | *Summative Value* |
| Clear | **Predictable** | True |
| Comprehensive | Reproducible | |
| Consistent | Rigorous | |
| Generalizable | Systematic | |
| Interpretable | Transparent | |
| | **Predictive** | |
| **Non-rationalist Epistemic Values** | | |
| Authentic | Engaging | *Summative Values* |
| Brave ("speaking truth to power") | **Generative** | Ethical |
| Caring | Imaginative | Beautiful |
| Creative | Nuanced | Rich (Generative, Suggestive) |
| Elegant | Playful | |
| Engaged | Responsible | |

*Figure 1*. Examples of epistemic values ordered as a structured set. "Predictive" is an example of a trickster value behaving undecidably in the structure. (Coyote drawing generated by ChatGPT.)

To borrow from structural anthropology, we might say that ordering epistemic values in this way is not unlike Claude Lévi-Strauss showing how peoples parse the world between nature and culture, or the raw and cooked. Epistemic values are *myths of knowledge*, and their apparent oppositions (accurate versus inaccurate, rigorous versus uncontrolled, and so on) invite us to look for the in-between tricksters—the coyotes and crows of Native American myth[18] or, in the context of AI, the hallucinations—representing undecidable value cruxes, which in the last analysis are what intellectual structures like this are ultimately hard at work trying to process. The term *predictive* is such a trickster value of AI. It fits ambiguously somewhere between values such as *generative* (able to connote *creative* or *rich*) and its opposite, *predictable*. Whenever we encounter the term *predictive*, especially anywhere in the neighborhood of the word *stochastic*, we should hear a coyote howling in the hills.

Nor is all this just in theory. Part of the point of making an anthropological comparison is to emphasize that structures of epistemic value affect social, political, and economic structures. Epistemic values articulate—e.g., put out the job calls for—today's knowledge work. The values *accurate*, *objective*, and *thorough* might be the job description of a good editor. Add *ethical* and *brave* to put out the job call for an investigative journalist. And combine *creative*, *imaginative*, *original*, and *eloquent* to hire script writers, artists, designers, and other creatives. How clusters of epistemic virtues match up with jobs is constantly evolving. Society will soon need to decide which values associated with "good" knowledge work can be offloaded to AI so that the value spectrum within which humans can uniquely be rewarded narrows or, more ideally, shifts.

[18] On Native American trickster figures, including crow/raven and coyote, see Ballinger.

Really, though, a zero-sum understanding of *human vs. machine* may not be right for calibrating AI's sociotechnical value. To tease a thought: perhaps there can be new values in the spectrum that mix the colors of human and machine.

Thus consider creativity.[19] AI is already affecting creativity work in ways that might have been foreseen from Margaret A. Boden's *The Creative Mind: Myths and Mechanisms* (1990; revised edition 2004). In recent articles such as Giorgio Franceschelli and Mirco Musolesi's "On the Creativity of Large Language Models," Boden's book has become a touchstone of thought on AI and creativity.[20] While Boden herself preceded the arrival of generative AI, her searching reflections on early examples of autonomous machine generativity (such as Harold Cohen's algorithmic artist, Aaron) are now more than ever relevant. Boden identifies three primary kinds of creativity. One is *combinatory* ("making unfamiliar combinations of familiar ideas"). The second is *exploratory* ("We can compare this with driving into the country…. [S]uppose … you drive off onto a smaller road …"). And the third is *transformative* (like "building a new road … a new technique, leading to new possibilities[,] or even re-routing the motorway"). The latter is highest on Boden's value scale: "The deepest cases of creativity," she says, "involve someone's thinking something which, with respect to the conceptual spaces in their minds, they *couldn't* have thought before" (Boden 3–6).

Extrapolating to current AI suggests the following. Generative AI will swallow up jobs devoted to combinatory creativity. ChatGPT, for example, is nothing if not a maker of "unfamiliar combinations of familiar ideas." Exploratory creativity (prospecting unfamiliar "smaller roads" in a pre-defined area) will likely be a partnership of humans with generative AI, though agentic AI with larger "context windows" and better "harnesses" to sustain long-running research tasks will become increasingly autonomous in exploratory creativity.[21] The highest rewards, however, will presumably go to humans who can be *transformatively* creative in ways still usually thought to be outside the reach of AI—at least until "artificial general intelligence" (AGI) arises if ever it can. AGI may well be the ultimate myth—the greatest trickster—in the AI story: not "stochastic parrot" (Bender et al.), as it were, but crow.

Yet to continue teasing my thought about wider spectra: how fixed are Boden's three colors of creativity anyway? Might AI extend the spectrum to new kinds of transformation for which the term *creativity* itself will no longer be adequate? After all, the very word *transformers* now describes an essential technology by which AI neural networks attend holistically rather than sequentially to context to enable modes of AI creativity for which we currently only have the placeholder term *generative*. Beyond just applying known epistemic values to AI, therefore (effectively, making them epistemic benchmarks analogous to technical benchmarks), an important project for the future will be to watch for evolving epistemic value assemblages--new value terms arising around AI, defamiliarizations of existing terms, and inventive combinatorics of such terms. The goal is both fresh positive appraisal, with less Silicon Valley and Wall Street hype, and, equally important, fresh negative critique, with less rote dystopianism.

On the positive side of the values ledger, consider the following articles in my corpora, which appreciate AI for new modes of creativity. Louise Vigeant affirms Boden's analysis of creativity (46-

---

[19] For my previous discussions of creativity in the context of postindustrial "creative destruction," see my *Laws of Cool* 317–71 and "Thinking Destruction."

[20] My thanks to Richard Jean So for bringing this article to my attention.

[21] For an explanation of AI "context windows," see Bergmann. On AI "harnesses," see Salesforce; and Böckeler.

48), but argues that generative AI advances a fused epistemic value of "argumentative creativity" that enmeshes "critical thinking skills, especially analysis and evaluation … with creativity (45). "Creativity lies in the combination of skills and abilities between human and technology, not in either alone" (61). Laavanya Ramaul et al. discuss new mixes of "conversational" and "creational affordances" in AI. And Jonathan Barlow and Jennifer Holt in their "Attention (to Virtuosity) Is All You Need" propose that AI hallucinations should be valued for creative "virtuosity." "Do we really desire an AI so constrained by rules, boundaries, and checkpoints," they ask, "that it could not achieve the novel and imaginative insights/foresights that are characteristic of intellectual virtuosity?" (7 of 11).[22]

The rehabilitation of AI hallucination attempted by Barlow and Holt is part of a wider recent effort to reevaluate hallucination positively. Inspired by the research method of "critical fabulation," for example, a group that includes Peiqi Sui, Eamon Duede, Hoyt Long, Richard Jean So, and Sophie Wu argue in preprints (postdating my corpora of 2024 articles) that AI hallucination is a form of *confabulation* supporting narrativizing functions that are crucial for making sense of things when information or context is missing:

> we attempt to broaden the concept of hallucination and argue that hallucination is closer in kind to the concept of "confabulation" (Sui, Duede, Wu, et al. 2).
>
> we define confabulation ... as a latent narrative impulse to generate more substantive and coherent outputs—a characteristic of LLM textual outputs that closely mirrors the human predisposition to storytelling as a cognitive resource for sensemaking. Specifically, confabulation is a narrative impulse to schematize the information at hand into self-consistent stories, even if there might not be enough available details to do so…. (Sui, Duede, Wu, et al. 3).
>
> In the academic field of African American studies, [Saidiya] Hartman introduces critical fabulation, the practice of using speculative storytelling to rectify omissions in historical archives due to social and political inequality, which has since become a field-defining methodology adopted in a wide range of humanistic historical case studies (Sui, Duede, Long, et al. 2).
>
> We argue that confabulation's narrative-rich properties should not be viewed as a flaw but a hallmark for LLM alignment with a well-established human tendency to use narratives as a versatile tool for persuasion, identity construction, and social negotiation (Sui, Duede, Wu, et al. 5).[23]

The negative side of the AI values ledger is also intriguing. The flip side of AI optimism, of course, is typically pessimism. To allude to the Terminator movies, if one is not a utopian who trusts that AI is all blue sky, then one is a dystopian who thinks it is Skynet. Such dystopianism can sometimes be too predictable, which in the humanities means it is open to the charge that it is stuck in a rut of "critique."[24] Yet dystopianism easily exceeds the rote wherever in fiction or criticism it is let loose to spawn irrepressibly gothic, ironic, or tragicomic golem creatures of the imagination that take on a life of their own. Some of the most prevalent, recent epistemic values invented to stigmatize bad AI are such golems, whose shared trait is that they seem to be directly antithetical to creativity. The rogues in this gallery include *homogeneity*, *monoculture*, and *slop*,[25] as well as, more impolitely, *enshittification*. The latter is Cory Doctorow's influential term for how online platforms become polluted with low-quality material through a lifecycle in which they first offer value for the end user, then prioritize value for sellers and advertisers, and finally shovel out as much slop or (in keeping with the trope) crap as possible to boost profits for their shareholders. As in Tim O'Reilly's "Where Is AI on the Enshittification Curve?," the term *enshittification* and its scatological synonyms are now increasingly applied to AI as both an enabler and peak example. To play on the older computational trope of *garbage in, garbage out*, perhaps AGI will be not *artificial general intelligence* but *artificial garbage in-and-out*.

---

[22] Vigeant and Barlow, and Holt are in my WoS AHCI corpus; Ramaul et al. is in my WoS Core corpus.

[23] These representative quotations omit the empirical evidence that the authors adduce. Thanks to Richard Jean So for pointing me to these preprints.

[24] E.g., see Rita Felski's *The Limits of Critique*.

To introduce an anachronistic, but uncannily prescient, allusion: we might think of AI *homogeneity*, *monoculture*, *slop*, and all such machine-learning *shit* as a modern version of Rabelaisian word vomit and logorrhea—that is, language regurgitated and reshat in an endless loop to produce (euphemizing) *word salad*. In this conceit, Gargantua and Pantagruel were the original large (giant!) language models spewing innumerable word lists such as the following from Gargantua: "I wiped my tail with a hen, with a cock, with a pullet, with a calf's skin, with a hare, with a pigeon, with a cormorant, with an attorney's bag, with a montero, with a coif, with a falconer's lure…" (Rabelais). Yet, and this is the point of the allusion, Rabelais's prose monsters also turn language and everything else processed through their guts (neural networks today) into *fecundity*. My allusion to Rabelais hints at the trickster phenomenon by which AI homogeneity, monoculture, slop, and similar crap can unpredictably morph from negative to positive epistemic value. A 2026 paper by Kommers et al. titled "Why Slop Matters" sets out to establish the research framework for identifying such value. Speculating on why people have an "appetite" for "consuming" AI slop, which their paper initially compares to "off-cuts of intelligence" (like offal served at Gargantua's table) Kommers et al. inquire into slop's "social function" ("AI Slop does *something*") and also its uniquely sloppy "aesthetic value," which may yet emerge in the way that such earlier found-values as "kitsch," "camp," and "pastiche" emerged from "low culture" (3).

Indeed, *emergence* may be the very germ, polluted yet animating, of what I termed *fecundity* above. To dispense with the need for allusion: we should recognize that the great golem lurking in the background is that teeming, fertile trickster of modern physics and information theory: *entropy*. The very name of one our most capable current LLMs, Anthropic's Claude, after all, celebrates Claude Shannon's work on information as entropy. On the one hand, AI slop is the entropy of language: a homogeneous slurry that—in the manner of Shannon's famous dictum that the "semantic aspects of communication are irrelevant to the engineering problem" (Shannon and Weaver 31)—is only tangentially, if at all, meaningful. But, on the other hand, entropy is a fecundity that surprises with the trick of *emergence*. A comparison here might be cellular automata, agent-based models, or boids (flocking or swarming algorithms). The fact that all the cells in a system like Conway's Game of Life are homogeneously the same, and follow a narrow set of rules in reaction to local neighbors, does not rule out the emergence from slop (the majority of end states of the Game of Life that are just static, random noise or nothing at all) of unpredictably complex structures and behaviors (Wikipedia).

To conclude, my argument is that we need to be fresher in the epistemic values we assign to AI, good or bad. We need to be receptive to, even while being careful about, new values and the sociotechnical structures shaping them.

---

[25] On AI "homogenization," "monoculture," and "slop," see e.g. Anderson et al.; Bommasani et al. 3–6, 152–53; Kleinberg and Raghavan; and Kommers et al. The last-mentioned article is interesting in my context because it reevaluates AI "slop" to find redeeming social value in it. (My thanks to Richard Jean So for pointing me to several of these works.)

To invite others to participate in exploring the evolving language of values about, and swirling around, AI, I present my online "Kit for Exploring Articles on AI Published in 2024" (https://alanyliu.org/citation/kit-for-exploring-articles-on-ai-published-in-2024/).[26] Included are my data models (topic models and word embeddings), interactive visualizations, in-progress list of found epistemic values, and documentation (see figure 2).[27]

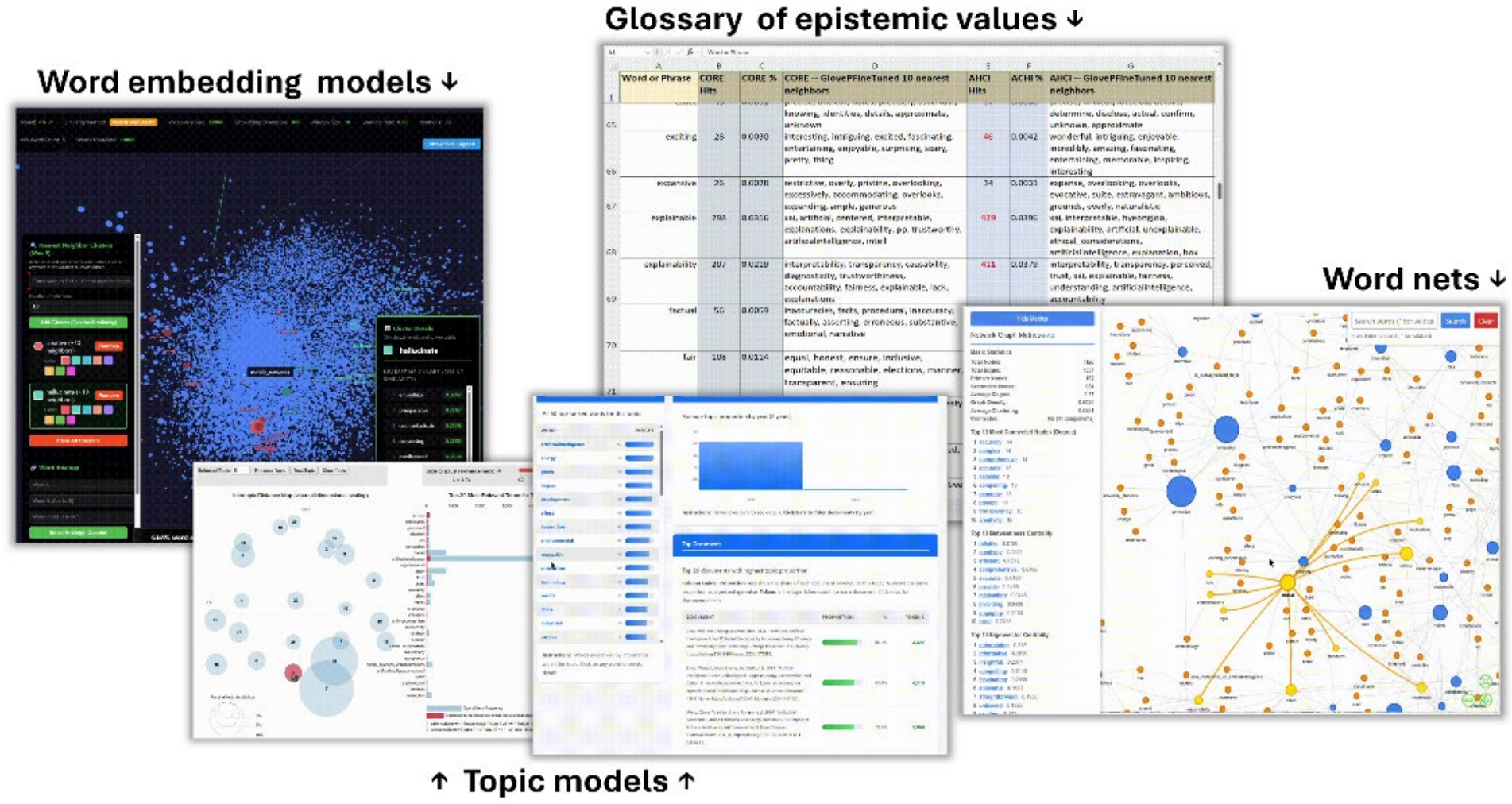

*Figure 2. Selected data and tools in "Kit for Exploring Articles on AI Published in 2024," including interactive visualizations of word embeddings, topic models, and word nets, and an in-progress glossary of epistemic values.*

How to use such a kit to observe epistemic values? My practice has been incremental, branching, and iterative. Start anywhere on a word associated with a known epistemic value. I will call such a *seed value*.[28] As an example, take *responsibly* as a seed value in the word-net of epistemic values I harvested from the embedding space for my WoS AHCI corpus (see figure 3). Then harvest related epistemic values by branching out to ngrams, collocates, or nearest word-embedding neighbors.[29] For instance, hop from *responsibly* to *honestly*, and then, somewhat surprisingly, to *creatively* a further hop away.

---

[26] For sustainability my kit can also be downloaded from a permanent deposit in the Zenodo open data repository at https://doi.org/10.5281/zenodo.19264354. I did not also put my kit in GitHub because some files exceeded GitHub's file size constraints.

[27] *AI use disclosure:* for making topic models and word embedding models, I used established machine-learning tools. As documented in my kit, for topic modeling I used Mallet (McCallum), and for word embedding I fine-tuned a pretrained GloVe word embedding (Pennington et al.) on my corpora. I also used established tools for displaying my topic models interactively: pyLDAvis (Mabey) and dfrbrowser (in a new DFR Brower 2 version by Kleinman). However, for displaying my word embeddings I used AI (Anthropic's Claude) to "vibe code" Python scripts for making standalone interactive HTML pages. (I then edited the scripts to customize further). Similarly, I used Claude to create scripts for making the interactive network graphs of my spreadsheet of harvested epistemic values and nearest embedding neighbors. I also used AI (ChatGPT and Claude in tandem) to help label topics in my topic models, though I edited some labels after familiarizing myself with the topics, their keywords, and their associated articles. For more detail about the making of my kit—including comments on the kit's limitations and constraints—see "readme" documents in the kit.

[28] Cf., Heuser and Le-Khac's term "seed word" in their study of "semantic cohorts" in nineteenth-century British fiction.

[29] For readers unfamiliar with word embedding, a good introductory explanation is by Barnard. Schmidt, and Heuser, provide introductions specifically for humanities scholars.

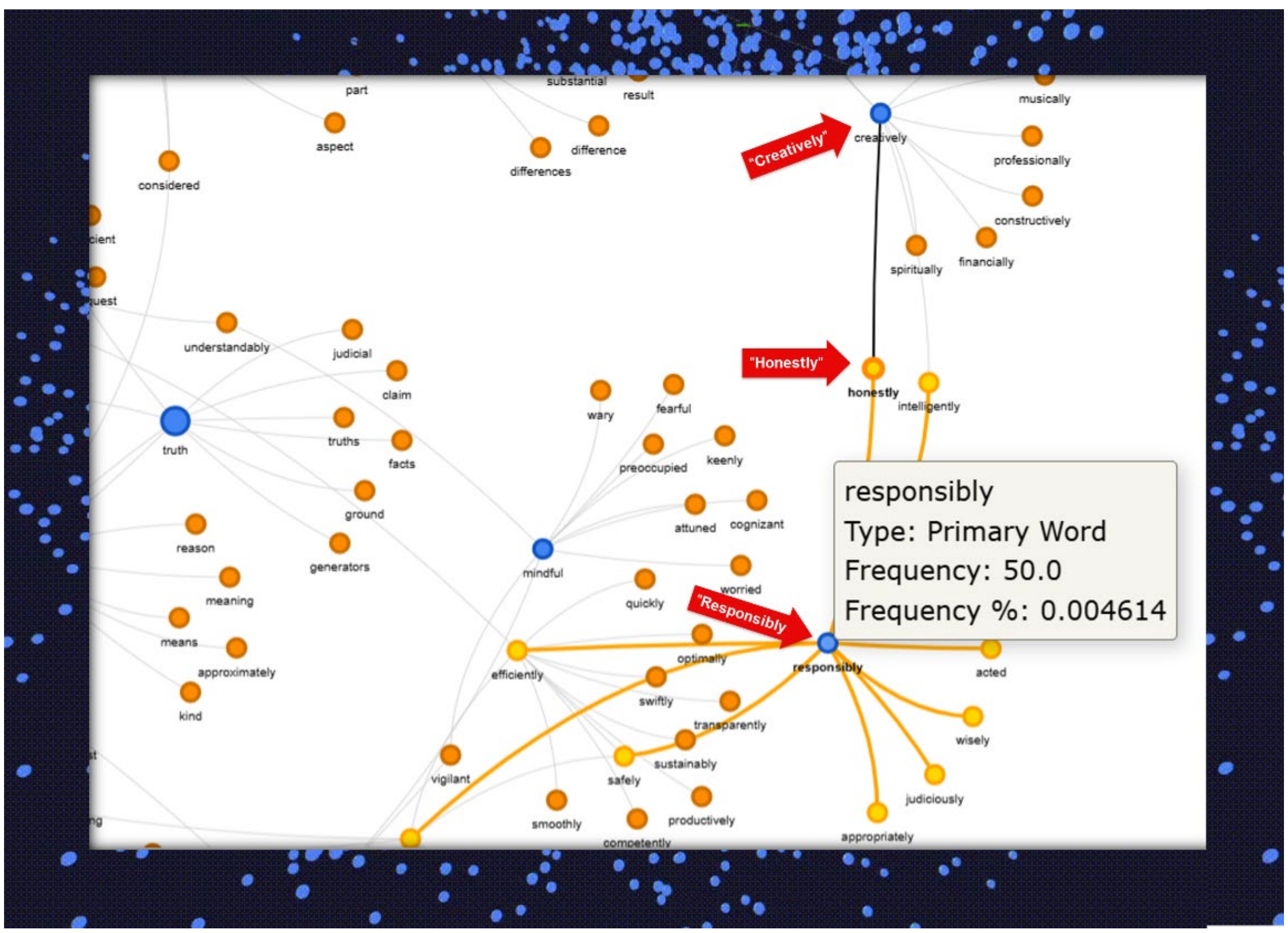

*Figure 3. Part of word-net graph of epistemic values (and nearest word embedding terms) in articles on AI published in 2024 in arts and humanities journals. Flagged here is the epistemic value "responsibly" and then, one and two hops away in the embedding space, "honestly" and "creatively."*

Next, repeat, using found values as new seeds. Very rapidly, this builds a word list like the "Glossary" in my kit. Adding to my glossary by making successive passes through the corpora with multiple tools—e.g., *AntConc* for corpus linguistics study (Anthony, *AntConc*), Mallet for topic modeling (McCallum), and GloVe for word embedding (Pennington et al.)—can then fill out the terminological map so that a picture emerges not just of specific domains and scales but of the larger epistemic structure.

In the future, linguistic mapping like this can be made more robust through more expert applied linguistics, "grounded theory" qualitative analysis, controlled vocabularies, and collaboration among domain experts.[30] Eventually, it should be possible to raise the analysis to a higher level to discover grouped assemblages of values like the hypothetical groupings in figure 4, which would represent epistemic norms.

[30] On qualitative analysis and grounded theory methods, see Zhang and Wildemuth.

| accurate, precise, efficient, reliable, unbiased, relevant | brave, bold, virtuous, passionate, principled | careful, cautious, prudent, judicious, reasoned, thoughtful | complex, nuanced, multifaceted, sophisticated, intricate |
| --- | --- | --- | --- |
| comprehensive, systematic, broad | creative, artistic, innovative, original, unique | critical, reflective, essential | engaging, engaged, interactive, effective |
| ethical, responsible, fair, honest, transparent, accountable, trustworthy | interpretable, explainable, generalizable | logical, predictable, coherent, rational, reasonable, | personal, private, individual |
| playful, fun, entertaining, enjoyable, humorous, expressive | replicable, reproducible, observable, interoperable | trustworthy, trusted, responsible, competent, secure, safe | valid, consistent, convergent, reliable |

*Figure 4*. Hypothetical groupings of epistemic values that may represent epistemic norms in discourse on AI.

Thus consider how the kit can be used to explore my specific case study in this essay: creativity as a value. The following are three observations made with the kit, each leading to an intriguing research question.

(1) Observing the words *creative*, *creativity*, and *creatively*,[31] we see that there is a notable difference between their frequency in my WoS Core and WoS AHCI corpora. While in WoS Core the first two words number around .050-.063% of total tokens, those words are far more frequent in WoS AHCI (*creative*, .144%; *creativity*, .083%).[32] This difference is not unexpected, of course, since WoS Core is dominated by healthcare, business, education, regulatory or public policy, and similar topics, while WoS AHCI is sourced from journals in the humanities and arts. Nor, inversely, is it surprising that WoS Core far exceeds WoS AHCI in mentions of epistemic values such as *innovation* that are standard proxies for creativity in the vocabulary of the postindustrial "innovation economy" (WoS Core: *innovation*, .246%; *innovative*, .039%; WoS AHCI: *innovation*, .079%; *innovative*, .024%).[33]

Such findings, though expected, set the agenda for further questions. One is the following. Given that terms such as *creative* and *innovative* occur in specific organizational and disciplinary research contexts, what *inter*organizational or *inter*disciplinary research is most promising for rethinking AI-related terms like *predictive* and *generative* so that they begin to absorb some of the meaning of *creative*—or, put another way, to shape new organizational and disciplinary contexts in which it becomes meaningful to say that AI is creative? The epistemic value of *predictive*, *generative*, and other words specific to generative AI are still malleable, and would benefit from escaping the vise of pre-existing knowledge work structures.

[31] I did not stem or lemmatize my corpora because in some cases only specific parts-of-speech or inflected forms of a word in a word family signify an epistemic value. (On word families, see note 17 above.)

[32] The spreadsheet glossary in my kit (folder 4) records the frequency of occurrences of each word in my WoS Core and WoS AHCI corpora as a raw number ("hits") and also—to normalize for the larger number of articles in the latter corpus—as a percentage of the total tokens in each corpus.

[33] J. P. Morgan's report titled "Innovation Economy Update H2 2025" (Candy and Harrison) is typical in using "innovation" as a metonymy for the overall economy. The report discusses AI in several sections.

(2) Observing *creative* as it occurs in ngrams, we see that the word modifies not just artificial intelligence (e.g., from my WoS Core corpus: "*creative AI* might initially help discover breakthrough innovations," Grewal et al. 874) but also many agents, actions, and their attributes in the total epistemic system around AI populated by words like *agency, applications, cognition, collaboration, content, design, expression, ideas, individuals, industries, process, skills, thinking, work,* and *writing*. The view of this epistemic system becomes even fuller when we observe *creative*'s collocates (e.g., *generate* and *generating* are among frequent collocates in WoS Core, while *god* stands out starkly among frequent collocates in WoS AHCI).[34] Word-embedding nearest neighbors (e.g., *ideas, thinking, article, material, work, generate, artistic*) add to the picture too.

Such exploration allows us to ask how linked complexes of AI, humans, and organizations *together* can be creative, which in the end is more meaningful than asking if AI is creative. In other words, exploration like this helps assess how the whole system of creativity in which generative AI plays an increasing role is expanding, contracting, or morphing in conjunction with sociocultural changes. Research questions for the future thus include: which areas of the total epistemic and social graph (glimpsed in such actual graphs as the word-embedding and word-net visualizations in folders 3 and 4 of my kit) will flex first under the influence of AI to connect with which other areas? Also, as suggested earlier, what will be the consequences for employment as jobs for creatives and others also flex to redistribute tasks among humans and machines?

(3) Finally, to bring this abbreviated demonstration to a close, we can begin to explore how the very idea of creativity might evolve beyond categories such as Boden's to undergo a phase shift into a new spectrum of creativity that may be called *generativity*. As noticed above, frequent collocates of *creative* in my corpora are *generate* and *generating*; and, correspondingly, word embeddings show that words in the *creativity* and *generativity* word families are near neighbors. But consider what happens if we then follow semantic trails outward from the *generative* word family. In my word embedding models (e.g., for my WoS AHCI corpus), one such trail leads to near neighbors like *capable* and *ability* (see figure 5).

[34] An AntConc analysis of my WoS AHCI corpus shows that among substantive words the top five collocates of *creative*, from most to least frequent, are: *artificial_intelligence*, *intended*, *thinking*, *writing*, *god*. (I omit stop words and also words like "commons" that are artifacts of journal articles because of specialized phrases like "creative commons.")

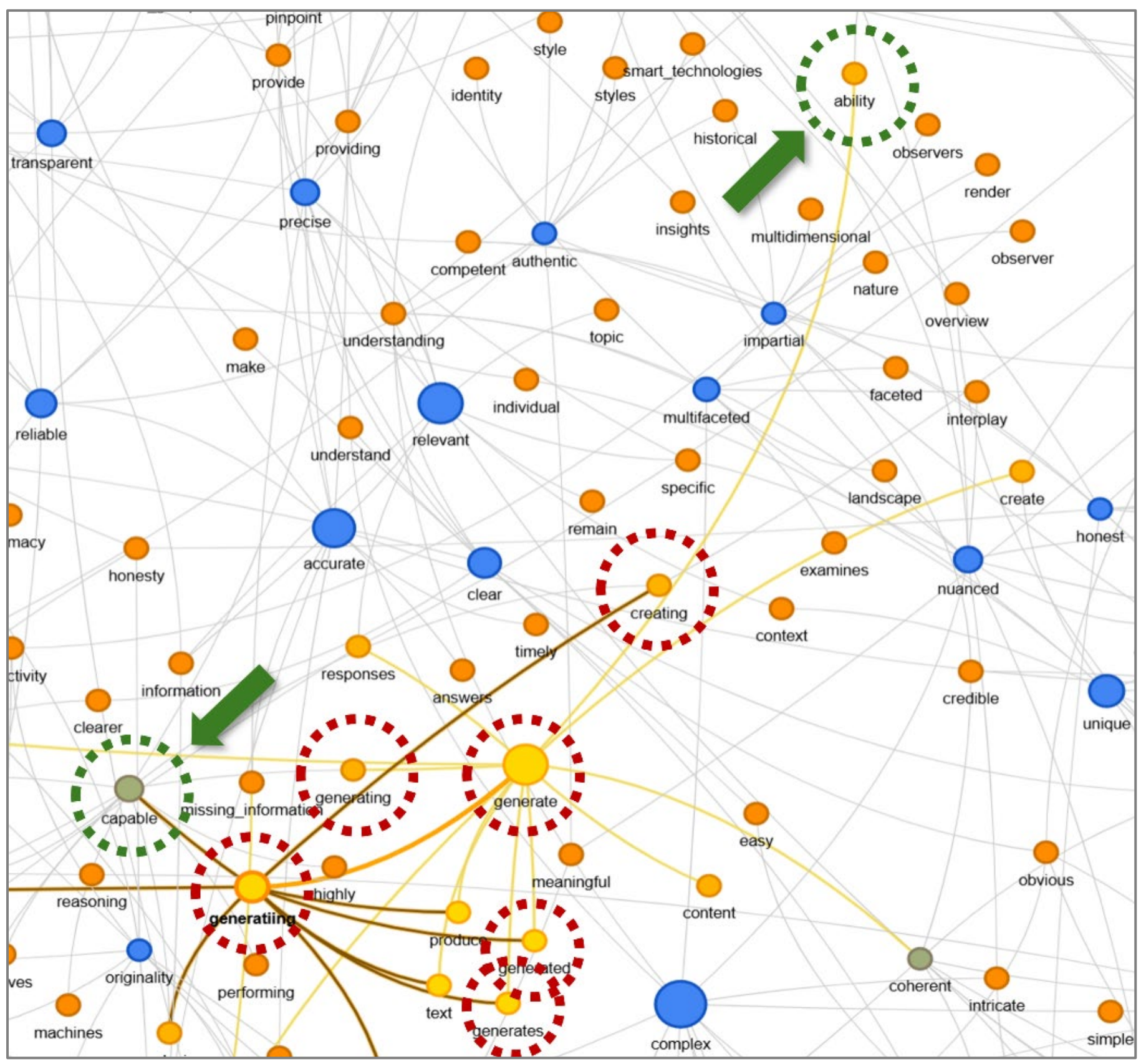

*Figure 5: Word-net graph of epistemic values and word-embedding neighbors for my WoS AHCI corpus. Terms in the "generative" family are close to "capable" and "ability" in the underlying embedding space.*

Seen one way, this registers a semantic difference from creativity. Creativity is genius or insight, we think, while generativity is just being capable in a workaday way. But seen another way, creativity, whose word family is *also* in the neighborhood, may itself be converging with workaday capability in the manner already hinted at by terms like *creatives* used to name a class of knowledge workers. It may be that a "broader understanding of generativity" allowing for new "creational affordances"—as one article in my WoS Core corpus puts it (Ramaul et al. 618)—can transform the idea of creativity so that in the future we will be able to speak of *generatives* (or some more elegant neologism) in the way we now speak of creatives.

My *Laws of Cool* book, I close by confessing, was in part an act of mourning for all my students destined to become knowledge workers in cubicles, or in today's possibly more imprisoning, because panoptically transparent, open-plan offices. They are secretly cool, yet all day they just write code or reports, or, beginning now, prompts for generative AIs to do so. Why should not creativity be democratized as a more inclusive ideal of generativity—one that workers and AI share, but one also for which, precisely because it *is* an ideal and not yet a norm, hard definitional and ideological battles will from this point on have to be fought?

**NOTE*:*** This article descends from a 2023 commencement speech I gave to English majors at my university (Liu, "What Is Good Writing in the Age of ChatGPT?"). I have greatly extended and altered my argument. The present essay also defers the original focus of that speech: good *language* in the age of AI. It would require another essay to take account of the relation between epistemology and linguistics that has been posed from the time of Enlightenment philosophy (e.g., John Locke) through modern ordinary-language, distributional-semantic, structural-linguistic, and deconstructive philosophy and linguistics. In such an essay, my present mapping of epistemic values would need to be extended to mapping value terms related to language, style, and voice.